\documentstyle[preprint,aps]{revtex}
\begin{document}
\def\sla#1{\rlap\slash #1}
\draft
\title{Covariant and Light-Front approaches to the
$\rho$-meson electromagnetic form-factors 
\footnote{Ref. Phys.Rev.{\bf C55} (1997) 2043.}}
\author{J.P.B.C. de Melo$^a$ and
T. Frederico$^b$}
\address{
$^a$ Instituto de F\'\i sica, Universidade de S\~ao Paulo, \\ 01498-970
S\~ao Paulo, S\~ao Paulo, Brazil. \\
$^b$ Departamento de F\'\i sica, ITA, 
Centro T\'ecnico Aeroespacial, \\
12.228-900 S\~ao Jos\'e dos Campos, S\~ao Paulo, Brazil.}
\date{\today}
\maketitle
\begin{abstract}
The  $\rho$-meson  electromagnetic form-factors
are calculated, both in a  covariant and  
light-front frameworks with constituent quarks.
The effect of the breakdown of rotational symmetry 
for the one-body current operator in the null-plane 
is investigated by  comparing calculations within light-front and
covariant approaches. This allows to choose
the appropriate light-front prescription, among the several ones, 
to best evaluate  the $\rho$-meson form-factors.
\end{abstract}
\narrowtext
\section{Introduction}

Since Dirac \cite{dirac}, it is known that the light-front hypersurface
given by $x^+=x^0+x^3=0$ (null-plane) is suitable for
defining the initial state of a relativistic system. 
Relativistic models with null-plane wave-functions have becoming 
widely used in particle phenomenology \cite{teren}.
They permit to calculate the  matrix elements of certain operators 
in a framework of a fixed number of constituents, while
mantaining a limited covariance, under transformations that keeps the
null-plane invariant \cite{wil90}.  As the generators of rotations 
around $x$ and $y$-axis do not belong to the stability group \cite{nam}, 
the covariance of a composite null-plane wave-function with a
fixed number of constituents can be broken.

This fact has consequences, for example, in the calculation of
the electromagnetic form-factors of a composite spin one particle
in the light-front\cite{frank}. It is well known  that, the
 $J^+(=J^0+J^3)$ component of the current, looses its 
rotational invariance and consequently violates the
angular condition in a light-front calculation \cite{frank,inna84}.
The light-front matrix elements are computed
with a null-plane wave-function\cite{inna84,chung88} 
in the Breit-frame, where the vector component of the
momentum transfer is along the $x$-direction.
If rotational symmetry around the $x$-axis
is valid,  then  $J^+_{zz}\ = \ J^+_{yy}$, where
the subscripts are
the polarizations of the spin one particle in the cartesian
instant-form spin basis \cite{frank93}. Such requirement is
the angular condition\cite{frank,inna84}. It can also be derived
in the front-form spin basis, using  general arguments
of parity and rotational invariance of $J^+$ \cite{keispol}.

The breakdown of rotational symmetry, implies that, it  
does not exist an unique way to extract electromagnetic form-factors
from the light-front matrix elements of $J^+$, for
composite systems with spin equal or higher than one.
Consequently, in the literature, there are
different  extraction schemes for spin one 
form-factors \cite{inna84,chung88,frank93,brod92}.

Recently, the issue of the breakdown of rotational covariance
for the one-body component of the $J^+$  current in a
light-front model, has been
discussed in the calculation of  $\rho$-meson form-factors with
constituent quarks \cite{keis94,inna95}.
In these works, it was stressed
the importance of relativistic effects related to the constituent mass
scale and the $\rho$-meson size. Such relativistic effects are smaller in a
test case of a S-wave deuteron system  and 
the violation of the angular condition is quite small \cite{fred91}.

The judgement of the different prescriptions for obtaining
electromagnetic form-factors  in the non-covariant light-front calculation,
which corresponds to use a specific $\rho$-meson 
 null-plane wave-function,  could be done in principle by  a
 comparision between the results of the non-covariant and
 covariant calculations, within the same model. 

It is our aim in this work to 
calculate the $\rho$-meson form-factors from the covariant Feynman
one-loop triangle-diagram  for the '+' component
of the current, in two ways, one corresponds to the covariant calculation
and the other to the non-covariant light-front calculation.
Below the covariant and the light-front calculations of the 
matrix elements of $J^+$ are explained.

The covariant calculation
corresponds to integrate directly the 
momentum loop in the standart variables in the
four-dimensional phase-space, integrating in  $k^0$ analytically.
In this case, the different prescriptions used to extract form-factors
from $J^+$, give identical results, since the matrix
elements  satisfy the angular condition. 

The non-covariant light-front  calculation corresponds to
the following procedure; the matrix elements of  $J^+$ 
 are obtained from the one-loop triangle diagram, which is
integrated analytically in the '-' component of
the loop momentum ($k^-=k^0-k^3$) in the Breit-frame with 
momentum transfer $q^+=0$, and numerically in
$k^+$ and $\vec k_\perp$. The Cauchy integration in $k^-$
takes into account only the propagator pole which corresponds to 
the forward propagation of the spectator particle, 
in the photon absorption process.
This in principle should be correct, as have been discussed
in the context of spin zero bound state particles \cite{tg}.
 However, we found numerically that the integration in $k^-$
 is not exact in the present case of spin 1 composite particle.  
The resulting matrix 
elements of $J^+$  do not satisfy the angular condition and
the form-factors now depend on  the prescription used.
Our numerical calculations 
show the difference between the two ways of performing the integrations. 
This may sound surprising, but we remind the reader that the $k^-$ 
integration is in general tricky and even in simple cases, 
taking into account only the forward propagating particle pole 
may give wrong results. The interested
reader can find examples in Ref.\cite{yam} with such problems.
The integration in $k^-$, taking into account only the above mentioned 
pole is equivalent to the use of a wave-function in the null-plane to
obtain form-factors in the Breit-frame\cite{tg}. 

To deal with a finite value for the triangle-diagram,
we introduce a covariant regulator, in a manner proposed in
Ref. \cite{shakin}. This allows the covariant integration
to be finite in one side and in the other side
the covariant regularization generates a null-plane $\rho$-meson 
quark wave-function \cite{shakin}.

We compare the covariant and  light-front
results for the $\rho$-meson form-factors, and
then, we are able to
point out the appropriate prescription to  evaluate the form-factors of
the $\rho$-meson with the null-plane wave-function. This work is relevant
in pratical situations, when one is faced with the problem of which
light-front prescription to use in the calculation
of the $\rho$-meson electromagnetic form-factors. In the present case,
 the parameters of the model are chosen such that it
reproduces to some extend the $\rho$-meson electromagnetic properties 
predicted by a realistic QCD inspired model \cite{inna95}.

The plan of the work is the following: in section II
is discussed  the different extraction schemes for obtaining
 the spin one electromagnetic  form-factors,
from $J^+$, and the notation is defined.
In section III, the matrix elements of $J^+$ are obtained from
 the Feynman triangle-diagram, and the covariant 
and  light-front calculations of the form-factors
are discussed. The
covariant regularization is shown to be related to the
null-plane wave-function.
In section IV, it is presented the numerical results for
the $\rho$-meson form-factors calculated in  
both frameworks and a summary of the main findings
are given.

\section{ electromagnetic current and form-factors}

The general expression of the electromagnetic current of a
 spin-one particle has the form \cite{frank} :
 
\begin{eqnarray}
J_{\alpha \beta}^{\mu}=[F_1(q^2)g_{\alpha \beta}-F_2(q^2)
\frac{q_{\alpha}q_{\beta}}{2 m_\rho^2}] P^\mu-F_3(q^2)(q_\alpha g_\beta^\mu-
q_\beta g_\alpha^\mu) \ ,
\label{eq:curr1}
\end{eqnarray}
where, $m_\rho$ is the  $\rho$-meson mass, $q^\mu$ is the 
momentum transfer and $P^\mu$ is the sum of the initial and final
momentum.

We  write the matrix elements of the $J^+$ component of the current
in the instant-form spin basis,  given by Eq. \ref{eq:curr1} 
in the Breit-frame, where $q^\mu=(0,q_x,0,0)$. The
 instant-form cartesian polarization four-vectors are given by:
\begin{eqnarray}
\epsilon^\mu_x=(-\sqrt{\eta},\sqrt{1+\eta},0,0) \ , \
\epsilon^\mu_y =(0,0,1,0) \ , \
\epsilon^\mu_z = (0,0,0,1) \ , \
\label{eq:pol} 
\end{eqnarray}
for the initial $\rho$-meson polarization states and,
\begin{eqnarray}
\epsilon_x^{'\mu}= (\sqrt{\eta},\sqrt{1+\eta},0,0) \ , \
\epsilon_y^{'\mu} = \epsilon_{y} \ , \
\epsilon_z^{'\mu}= \epsilon_{z} \ , \
\label{eq:polp}
\end{eqnarray}
for the final polarization states; $\eta\ = \ q^2/4 m_\rho^2 $.
The $\rho$-meson four-momentum  in the
Breit-frame are, 
$p^\mu_i=(p^0,-q_x/2,0,0)$ for the initial state, 
and  $p^\mu_f=(p^0,q_x/2,0,0)$for the final state;
  $p^0=m_\rho \sqrt{1+\eta} $.

We  use the spherical polarization vectors,
and in the instant-form spin basis are given by:
\begin{eqnarray}
\epsilon_{+-}=-(+) \frac{\epsilon_{x}+(-)\epsilon_{y}} {\sqrt{2}}
\ and \
\epsilon_{0}=\epsilon_{z} \ .
\label{eq:sppol}
\end{eqnarray}
The  "+" component of the electromagnetic current in the instant-form
spin basis is written as:
\begin{eqnarray}
J^{+}   =  \frac{1}{2}  \left( \begin{array}{ccc} 
J_{xx}^{+}+J^+_{yy}&  -\sqrt{2} J^+_{zx}    &  J_{yy}^{+}-J^+_{xx} \\
\sqrt{2} J^+_{zx}  &  2 J^+_{zz}  &  -\sqrt{2} J^+_{zx}          \\
J^+_{yy}-J^+_{xx}  &     \sqrt{2} J^+_{zx}   &  J_{xx}^{+}+J^+_{yy}  \\
\end{array} \right) \ ,
\label{eq:inst}
\end{eqnarray}
where the spin projections are in the following order $m=(+,0,-)$.
The  first  and second
subscripts of the current means the polarizations of the final
 and initial states, respectively.

The matrix elements of "+" component of the current in the
instant-form spin basis, Eq. \ref{eq:inst}, are related to
the matrix elements in the front-form spin basis.
For notational convenience, we use $I^+$, to
express the  matrix elements in the front-form spin basis. 
The unitary transformation
between these spin-basis is the Melosh rotation \cite{keispol}
 (Appendix). The general
form of the  "+" component of the current in the front-form
spin basis is written as \cite{frank93}

\begin{eqnarray}
I^+ =  \left( \begin{array}{ccc}
I_{11}^{+}  & I_{10}^{+}  & I_{1-1}^{+} \\
  - I_{10}^{+}  & I_{00}^{+}  &  I_{10}^{+}         \\
I_{1-1}^{+}  &   - I_{10}^{+}   & I_{11}^{+}  \\
\end{array}  \right) \ .
\label{eq:ifront}
\end{eqnarray}

We express the matrix elements in the front-form spin basis in terms of
 matrix elements of the current
in the instant-form spin basis, by using the results
of the Appendix,
\begin{eqnarray}
I^{+}_{11}&=&\frac{J^{+}_{xx}+(1+\eta) J^{+}_{yy}-
\eta J^{+}_{zz}+2 \sqrt{ \eta} J^{+}_{zx}}{2 (1+\eta)}
\nonumber \\
I^{+}_{10}&=&\frac{\sqrt{2 \eta} J^{+}_{xx}+\sqrt{2 \eta} J^{+}_{zz}
+\sqrt{2} (\eta-1) J^{+}_{zx}}{2(1+\eta)}
\nonumber \\
I^{+}_{1-1}&=&\frac{-J^{+}_{xx}+(1+\eta) J^{+}_{yy}+
\eta J^{+}_{zz}-2 \sqrt{\eta} J^{+}_{zx}}{2 (1+\eta)}
\nonumber \\
I^{+}_{00}&=&\frac{-\eta J^{+}_{xx}+J^{+}_{zz}+2 \sqrt{\eta} J^{+}_{zx}}
{(1+\eta)} \ .
\label{eq:ifront1}
\end{eqnarray}

The angular condition $J^+_{zz}\ = \ J^+_{yy}$ can be 
written in  front-form spin basis (see Appendix),
giving its usual form  \cite{inna84,keispol}
\begin{eqnarray}
\Delta(q^2)=(1+2 \eta) I^{+}_{11}+I^{+}_{1-1}-\sqrt{8 \eta} I^{+}_{10} -
I^{+}_{00} = 0
\ .
\label{eq:ang}
\end{eqnarray}

In general, the light-front impulse approximation to the electromagnetic
current does not satisfy such condition \cite{inna84,keis94,inna95,fred91},
this fact led to different extraction schemes of the form-factors
from the matrix elements of the current
\cite{inna84,chung88,frank93,brod92,inna95}.
Let us review the prescriptions existing in the literature for
calculating the form-factors
for  spin-one particle, from the matrix elements $I_{m'm}^{+}$.

The charge, $G_0$, magnetic, $G_1$, and quadrupole, $G_2$, form-factors
are obtained from linear combinations of the covariant
form-factors, $F_1$, $F_2$ and $F_3$ \cite{frank93}, see Appendix. 
Below, we give the different prescriptions for obtaining
the form-factors, which are also written in terms of matrix elements
in the
cartesian instant-form spin-basis. Such spin basis is used because
it facilitates the algebraic manipulations of the covariant amplitude
for the photon absorption process, and
 it is completly equivalent to the front-form spin basis.

In reference \cite{inna84}, the elimination of the 
matrix element $I^{+}_{00}$, gives
the following prescription to calculate the form-factors:
\begin{eqnarray}
G_0^{GK}&=&\frac{1}{3}[(3-2 \eta) I^{+}_{11}+
2 \sqrt{2 \eta} I^{+}_{10} +  I^{+}_{1-1}]
=\frac{1}{3}[J_{xx}^{+}+ 2 J_{yy}^{+}-\eta  J_{yy}^{+} 
+ \eta  J_{zz}^{+}]
\nonumber \\
G_1^{GK}&=&2 [I^{+}_{11}-\frac{1}{ \sqrt{2 \eta}} I^{+}_{10}]
=J_{yy}^{+} -  J_{zz}^{+}+\frac{J_{zx}^{+}}{\sqrt{\eta}}
\nonumber \\
G_2^{GK}&=&\frac{2 \sqrt{2}}{3}[- \eta I^{+}_{11}+
\sqrt{2 \eta} I^{+}_{10} -  I^{+}_{1-1}]
=\frac{\sqrt{2}}{3}[J_{xx}^{+}+J_{yy}^{+} (-1-\eta)
+\eta  J_{zz}^{+}] \ .
\end{eqnarray}

In Ref.\cite{chung88}, they have obtained
\begin{eqnarray}
G_0^{CCKP}&=&\frac{1}{3 (1+\eta)}
[(\frac{3}{2}-\eta) (I^{+}_{11}+I^{+}_{00})
+5 \sqrt{2 \eta} I^{+}_{10}+(2 \eta - \frac{1}{2}) I^{+}_{1-1}]
\nonumber \\
&=&\frac{1}{6}[2 J_{xx}^{+} + J_{yy}^{+} + 3 J_{zz}^{+}]
\nonumber \\
G_1^{CCKP}&=&\frac{1}{(1+\eta)}[I^{+}_{11}+I^{+}_{00}
-I^{+}_{1-1}  - \frac{2 (1-\eta)}{\sqrt{2\eta}} I^{+}_{10}]
=\frac{J_{zx}^{+}}{\sqrt{\eta}}
\nonumber \\
G_2^{CCKP}&=&\frac{\sqrt{2}}{3 (1+\eta)}[- \eta I^{+}_{11} 
-\eta I^{+}_{00}
+2 \sqrt{2 \eta} I^{+}_{10}- (\eta + 2) I^{+}_{1-1}]
=\frac{\sqrt{2}}{3} [J_{xx}^{+}-J_{yy}^{+}]
\end{eqnarray}

The prescrition of Brodsky and Hiller \cite{brod92}, to obtain the 
form-factors is:
\begin{eqnarray}
G_0^{BH}&=&\frac{1}{3(1+\eta)}[(3-2 \eta) I^{+}_{00}+
8 \sqrt{2 \eta} I^{+}_{10}+2 (2 \eta -1) I^{+}_{1-1}]
\nonumber \\
&=&\frac{1}{3 (1+2 \eta)}[J_{xx}^{+} (1+2 \eta)+ J_{yy}^{+}(2 \eta-1) 
+ J_{zz}^{+}(3+2 \eta)]
\nonumber \\
G_1^{BH}&=&\frac{2}{(1+2 \eta)}[I^{+}_{00}
-I^{+}_{1-1}+\frac{(2 \eta -1)}{\sqrt{2 \eta}} I^{+}_{10}]
\nonumber \\
&=&\frac{1}{(1+2 \eta)}[\frac{J_{zx}^{+}}{\sqrt{\eta}}
 (1+2 \eta)- J_{yy}^{+} +  J_{zz}^{+}]
\nonumber \\
G_2^{BH}&=&\frac{2 \sqrt{2}}{3 (1+ 2 \eta)}[\sqrt{2 \eta} I^{+}_{10}
-\eta I^{+}_{00} -( \eta+1) I^{+}_{1-1}]
\nonumber \\
&=&\frac{ \sqrt{2}}{3 (1+2 \eta)}
[J_{xx}^{+} (1+2 \eta)- J_{yy}^{+}(1+ \eta) 
- \eta J_{zz}^{+}]
\end{eqnarray}

According to Ref.\cite{frank93},  the electromagnetic form-factors,
are obtained from the 
 matrix elements $J^+_{xx}$, $J^+_{zx}$ and $J^+_{yy}$,
\begin{eqnarray}
G_0^{FFS}&=&\frac{1}{3 (1+\eta)}[(2 \eta+3) I^{+}_{11}+2 \sqrt{2 \eta} 
I^{+}_{10} -\eta I^{+}_{00}+(2 \eta +1) I^{+}_{1-1}]
=\frac{1}{3}[J^{+}_{xx}+2 J^{+}_{yy}]
\nonumber \\
G_1^{FFS}&=&G_1^{CCKP}
\nonumber \\
G_2^{FFS}&=&G_2^{CCKP}
\ .
\label{eq:g}
\end{eqnarray}

The low-energy $\rho$-meson observables,
mean square radius, magnetic moment and
quadrupole moment, are given by  \cite{chung88},
\begin{eqnarray}
<r^2>=\lim_{q^2 \rightarrow 0} \frac{6 (G_0(q^2)-1)}{q^2} \ , \
\mu=\lim_{q^2 \rightarrow 0}G_1(q^2)  \ , \
Q_2=\lim_{q^2 \rightarrow 0} 3\sqrt{2}\frac{G_2(q^2)}{q^2} \ ,
\end{eqnarray}
respectively.

\section{Covariant and light-front currents}

The $\rho$-meson  electromagnetic form-factors are  obtained
in the impulse approximation. It includes only one-body
 current operator, and the amplitude for
the photon absorption is given by the Feynman triangle-diagram,
with the photon leg attached to one of the quarks.
We compute only the "good" component of the
current ($J^+$), which is diagonal in the null-plane
Fock-state. The pair creation diagram
is in principle supressed for $J^+$  \cite{dash}.

The spinor structure of the $\rho - q \overline q $  vertex,
 is written in the following form, 
\begin{equation}
\Gamma^\mu (k,k') = \gamma^\mu -\frac{m_\rho}{2}
 \frac{k^\mu+k'^\mu}{ p.k + m_\rho m -\imath \epsilon}  \ .
\label{eq:rhov}
\end{equation}  
where, the $\rho$-meson is on-mass-shell, and its four momentum
is $p^\mu \ = \ k^\mu \ - \  k'^\mu$, the quark momenta
are given by $k^\mu$ and $k'^\mu$, and their mass by $m$.
Eq. \ref{eq:rhov}
reduces to the vertex given in Ref.\cite{ja90}
for a on-mass-shell quark.
 This vertex corresponds
to a relative S-state  quark-antiquark wave-function
\cite{frank,ja90}.
Above, we wrote
the spinor structure of the vertex. The complete null-plane 
wave-function comes from the regularization factor and the denominator
of the propagator, as it will be clear in the following.

The impulse approximation to $J^+$, is given by the  Feynman
triangle-diagram, and we assume the constituent quark as a Dirac pointlike
particle,
\begin{eqnarray}
J^+_{ji}&=&\imath  \int\frac{d^4k}{(2\pi)^4}
 \frac{Tr[\epsilon^{'\alpha}_j \Gamma_{\alpha}(k,k-p_f)
(\sla{k}-\sla{p_f} +m) \gamma^{+} 
(\sla{k}-\sla{p_i}+m) \epsilon^\beta_i \Gamma_{\beta}(k,k-p_i)
(\sla{k}+m)]}
{((k-p_i)^2 - m^2+\imath\epsilon) 
(k^2 - m^2+\imath \epsilon)
((k-p_f)^2 - m^2+\imath \epsilon)}
\nonumber \\ & &\times \Lambda(k,p_f)\Lambda(k,p_i) \ ,
\label{eq:tria}
\end{eqnarray}
where $J^+_{ji}$ is written in the cartesian instant-form 
spin basis, and $\epsilon^{'\alpha}_j$ is the
 final polarization four-vector
(Eq.\ref{eq:polp}) and $\epsilon^\beta_i$ is the initial
four-vector polarization  (Eq.\ref{eq:pol}), the subscripts $ i $ and $j$
stand for $ x$, $y$ and $z$.

The regularization function,
\begin{eqnarray}
\Lambda(k,p) =\frac{N}{((k-p)^2 - m^2_R+\imath\epsilon)^2 }\ ,
\label{eq:reg}
\end{eqnarray} 
was chosen to turn Eq.\ref{eq:tria} finite.
The special form of the regulator, allows to
identify a null-plane wave-function similar to the one
proposed for the pion in Ref.\cite{shakin}. They have
used a monopole form-factor, instead of a dipole. The normalization
factor $N$ is found by imposing $G_0(0)$=1 .

The covariant calculation of the form-factors,
is performed with  Eq. \ref{eq:tria}, which is analytically
 integrated in the $k^0$ complex-plane. The  integration over $\vec k$
is done numerically. The angular condition is satisfied exactly
by the covariant calculation, as it should be. 
This does not  remain true in the light-front calculation.
Also for
$q^2=0$, $J^+_{xx}(0)=J^+_{yy}(0)=J^+_{zz}(0)$. The matrix elements
of the current satisfy current conservation, 
$q^\mu J_\mu(q^2)\ = \ 0$, as we verified explicitly.

The light-front calculation, corresponds to integrate 
analytically in the complex-plane of $k^-$ variable \cite{tg}. 
In principle, the pair diagrams are not present with $q^+=0$. 
The pole which contributes to the integration is
\begin{eqnarray}
k^-=\frac{k^2_\perp+m^2-\imath\epsilon}{k^+} \ ,
\label{eq:oms}
\end{eqnarray}
for $ p^+\ > \ k^+\ >\ 0$, where $p^+=p^0$ is the energy of the
$\rho$-meson in the Breit-frame. This pole belongs to the lower
complex semi-plane, where no other pole is present. 
Eq. \ref{eq:oms}, is the on-mass-shell condition for the
spectator quark, in the process of photon absorption. 
However, the $J^+$
matrix elements do not satisfy the angular condition. Thus,
different prescriptions of calculating the form-factors
will give different results.
The analytical integration in $k^-$ for  the $\rho$-meson
including only the spectator particle pole is not exact 
as our numerical calculations have shown. In the next section,
we show our numerical results.

The null-plane wave-function of the $\rho$-meson appears
after the substitution of the on-mass-shell condition, 
Eq.\ref{eq:oms}, in the propagator of the quark that absorbs
the photon and in the corresponding regulator,
\begin{eqnarray}
\frac{1}{((k-p)^2 - m^2+\imath\epsilon)
((k-p)^2 - m^2_R+\imath\epsilon)^2 } =
\frac{1}{(1-x)}\frac{1}{(1-x)^2(m^2_\rho-M_0^2)(m^2_\rho- M^2_R)^2} \ ,
\label{eq:den}
\end{eqnarray}
where, $x=k^+/p^+$.  The free quark-antiquark   mass squared
is given by
\begin{eqnarray}
M^2_0= \frac{k^2_\perp+m^2}{x} 
+\frac{(\vec p-\vec k)^2_\perp+m^2}{1-x}-p_\perp^2 .
\end{eqnarray}
The function $M_R^2$ is given by
\begin{eqnarray}
M^2_R= \frac{k^2_\perp+m^2}{x} 
+\frac{(\vec p-\vec k)^2_\perp+m^2_R}{1-x}-p_\perp^2 \ .
\end{eqnarray}

The null-plane wave-function is obtained 
from the combination of denominators in Eq.\ref{eq:den} with the spinor  
structure of  the vertex, Eq.\ref{eq:rhov}.
We leave out the phase-space factor $1/(1-x)$. The resulting
expression is evaluated in the center of mass system, 
\begin{eqnarray}
\Phi_i(x,\vec k_\perp)=\frac{N^2}{(1-x)^2(m^2_\rho-M_0^2)
(m^2_\rho- M^2_R)^2} 
\vec \epsilon_i . [\vec \gamma -  \frac{\vec k}{\frac{M_0}{2}+ m}] \ ,
\label{eq:npwf}
\end{eqnarray}
the polarization state is given by $\vec \epsilon_i$. The wave-function
corresponds to a S-wave state \cite{ja90}.

\section{ Discussion}

The constituent quark model for the $\rho$-meson null-plane wave-function has
two parameters, the constituent quark mass, $m$, and the regulator mass,
$m_R$. The $\rho$-meson mass is 0.77 GeV. In this case, 
the composite wave-function corresponds to a  bound state, which
imposes a lower bound for the regulator and constituent quark masses,
such that 
$$ m \ > \ \frac{m_\rho}{2} \ , \ m_R + m \ > m_\rho\ . $$

The scale of the model is obtained by adjusting the parameters to get
a mean square radius of about 0.35 fm$^2$, 
and $G_2(q^2 \sim 5 GeV^2)\ \sim $ -0.25, 
as calculated in Ref.\cite{inna95}, with  point-like constituent quarks.
They used  a wave-function in the null-plane which is
dominated by one-gluon exchange  at short distances and linear 
confinement at large distances. 

In the non-relativistic limit, the quadrupole form-factor vanishes for
a S-state wave-function. The non-zero values of $G_2$
are a consequence of the relativistic
nature of the model, and for this reason we
consider it in the parameter fit.
We used the covariant calculation for the form-factors to get the
parameters $m=$ 0.43 GeV and $ m_R=$ 1.8 GeV.

The low-energy electromagnetic parameters, are calculated
using the different light-front prescriptions with the
light-front calculation of $J^+$ matrix elements and  are compared
with the covariant results. In Table I,
we show the values of $<r^2>$, $\mu$ and $Q_2$. The mean 
square radius, 
calculated in the light-front scheme has values at most
 10\% higher than the covariant result of 0.37 fm$^2$. The magnetic moment
obtained
in the covariant calculation is 2.19, which can be compared with
the non-relativistic value of 2. In Ref.\cite{inna95} they obtained 2.26.
The light-front calculations for the magnetic moment, 
give values with a spread of 15\% above
the covariant result. The quadrupole moment in the covariant
calculation is 0.052 fm$^2$, somewhat higher than the value quoted
in Ref.\cite{inna95}.
The light-front calculations are within 10\% to 15\% of the covariant result
 for the low-energy parameters.

In Fig. 1, we observe that the charge form-factor, $G_0$, 
is sensitive to the different light-front prescriptions.
The calculations show a zero placed around 3 GeV$^2$ consistent with 
Ref.\cite{inna95}. We found an increasing discrepancy 
among the several prescriptions and the covariant results, for
momentum transfers above the zero crossing. The (GK) prescription
gives results in  agreement with the covariant calculation, while the
(BH) results are about 30\% below at higher $q^2$.

The differences between the various light-front 
calculations for the magnetic form-factor and the
covariant results are not so  pronounced, as shown in Fig. 2.
At small momentum transfers the (FFS) prescription has a value
about 15\% higher than the covariant result, in agreement with the
results of Table I. 
In the momentum range considered, the (GK) prescription
is consistent with the covariant calculation.

The relativistic effects in the model are the origin of $G_2$,
and thus it is more sensitive to the difference between the light-front
 prescriptions. In Fig. 3,
the values of $G_2$ calculated in the light-front with
prescriptions given by (CCKP) and (BH) are 20\% lower than the
covariant result. The calculations with (GK) combination of the
currents, present the best consistency with the 
covariant results, among the four prescriptions tested.

We conclude that, in the scale of the $\rho$-meson  bound state, 
tunned by a parametrization which reproduces
the size and quadrupole form-factor, of an effective constituent
quark model, which embodies gluon exchange and confinement;
the prescription for a light-front calculation 
of the form-factors as given by
the work of Grach and Kondratyuk \cite{inna84} shows consistence
 with the covariant results. We have used
a vertex for the $\rho$-meson, that was amenable to covariant
integration and  reproduced to some extend
the size properties of a physically inspired null-plane wave-function.

\section*{Acknowledgements}
This work was supported by  Brazilian agencies CNPq,
CAPES and Probral/CAPES/DAAD.

\newpage

\appendix
\section{}
\setcounter{equation}{0}

The Melosh rotation for  spin 1 particle is given by:
\begin{eqnarray}
R_{M}   =    \left( \begin{array}{ccc} 
\frac{(1+\cos\theta)}{2}  & -\frac{\sin\theta}{\sqrt{2}} & 
\frac{(1-\cos\theta)}{2}  \\
\frac{\sin\theta}{\sqrt{2}}  & \cos\theta  & -\frac{\sin\theta}{\sqrt{2}}  \\
\frac{(1-\cos\theta)}{2} & \frac{\sin\theta}{\sqrt{2}} & 
\frac{(1+\cos\theta)}{2} \\
\end{array} \right) \ ,
\end{eqnarray} 
where $\cos\theta=(\sqrt{1+\eta})^{-1}$ and 
$\sin\theta=-\sqrt{\frac{\eta}{(1+\eta)}}$.

The matrix elements of the current in the instant-form spin basis $(J^+)$,
Eq.\ref{eq:inst}, and
in front-form spin basis $(I^+)$, Eq.\ref{eq:ifront}
are related by the Melosh rotation,
\begin{eqnarray}
R^\dagger_{M} I^{+}R^\dagger_{M}= J^{+} \ .
\end{eqnarray} 

The instant-form matrix elements are expressed in terms of the
light-front matrix elements as \cite{frank93}, using the above equation,
\begin{eqnarray}
J^{+}_{xx}&=&\frac{1}{1+\eta}[I^{+}_{11}+2 \sqrt{2 \eta} I^{+}_{10}-
\eta I^{+}_{00} - I^{+}_{1-1}]
\nonumber \\
J^{+}_{zx} &=& \frac{\sqrt{2}}{1+\eta} [\frac{\sqrt{2 \eta}} {2} I^{+}_{11}+
(\eta-1) I^{+}_{10}+\sqrt{\frac{\eta}{2}} I^{+}_{00} - 
\frac{\sqrt{2 \eta}}{2}I^{+}_{1-1}]
\nonumber \\
J^{+}_{yy}&=&I^{+}_{11}+I^{+}_{1-1}
\nonumber \\
J^{+}_{zz}&=&\frac{1}{1+\eta}[-\eta I^{+}_{11}
+2 \sqrt{2 \eta} I^{+}_{10}+ I^{+}_{00}+ \eta I^{+}_{1-1}]
\ .
\end{eqnarray}

The relations between the form-factors $G_0$, $G_1$ and $G_2$
and the covariant form-factors $F_1$, $ F_2$ and $ F_3$, are
given by:

\begin{eqnarray}
G_{0}&=&-\frac{2}{3}m_\rho\sqrt{1+\eta} [3F_{1}+2\eta (F_1+F_3+(1+\eta)F_2)]
\nonumber \\ 
G_{1}&=&2 m_\rho\sqrt{1+\eta}F_3
\nonumber \\
G_{2}&=&-4\frac{\sqrt{2}}{3}m_\rho \eta \sqrt{1+\eta}
[F_{1}+(1+\eta) F_2+F_3] 
\ .
\end{eqnarray}

\newpage

\begin{table}

\caption{ Results for the low-energy electromagnetic 
$\rho$-meson observables, for the covariant (COV)
and light-front calculations.
The light-front extraction schemes to obtain the form-factors 
are  given by Refs. (GK) [6], (CCKP) [7],  (FFS) [8] and
(BH) [10]. In the last column, the results of Ref. [12] are given.}
\vspace{1 cm}
\begin{tabular}{|c|c|c|c|c|c|c|} 
MODEL        &  COV  &   GK  & CCKP  &  BH   & FFS  & Ref.[12]\\ \hline 
$<r^2>(fm^2)$& 0.37  &  0.37 & 0.38  & 0.40  & 0.39 & 0.35 \\ 
$\mu$        & 2.14  &  2.19 & 2.17  & 2.15  & 2.48 & 2.26\\  
$Q_2(fm^{2})$& 0.052 & 0.050 & 0.051 & 0.051 & 0.058& 0.024\\ 
\end{tabular} 

\end{table}

\newpage

\begin{figure}[h]
\vspace{10.0cm}
\includegraphics{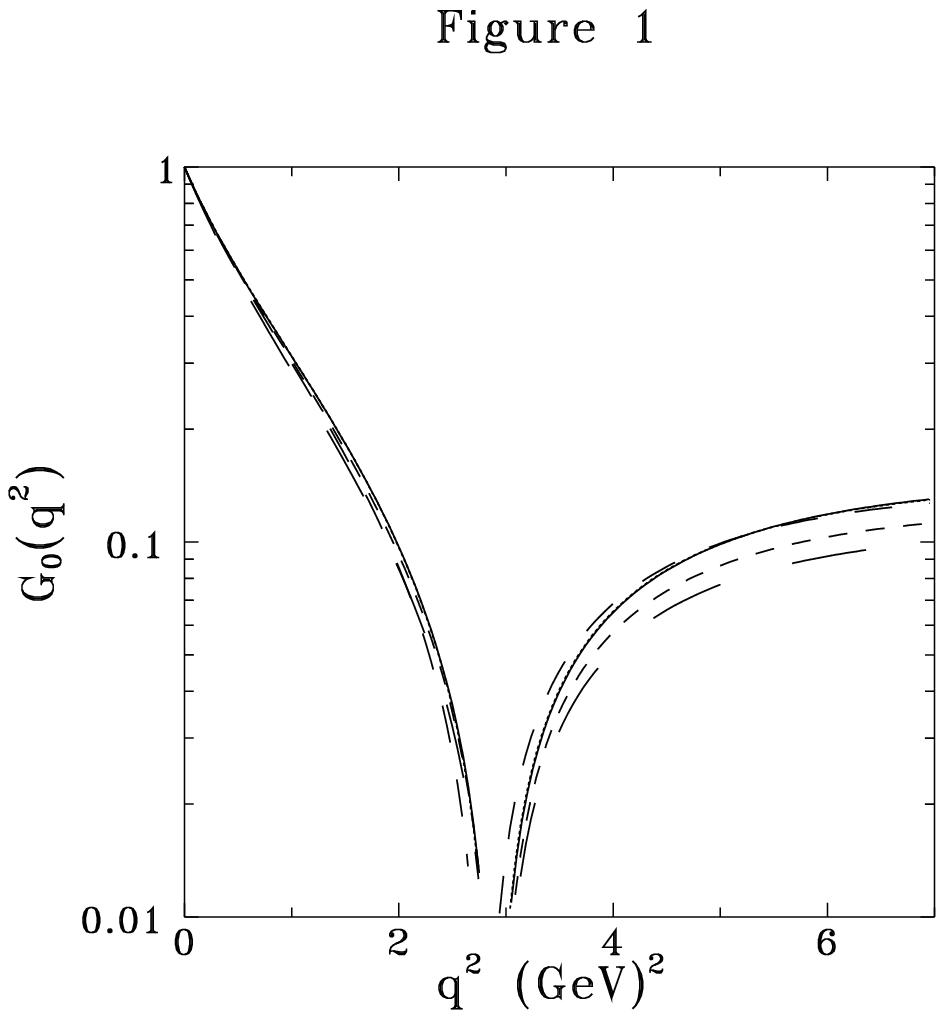}
\caption{ Charge form-factor $ G_0(q^2) $ 
for the $\rho$-meson as a function of $q^2$, calculated
with covariant and light-front schemes.
The solid line is the covariant calculation.
Results for the different light-front
extraction schemes, Ref.[6] (GK) (dotted line)
(it is not possible to distinguish from the covariant calculation),
Ref.[7] (CCKP) (short-dashed), Ref. [8]
(FFS) (dashed) and Ref.[10] (BH) (long-dashed).}
\label{fig1}
\end{figure}

\newpage
.

\begin{figure}[h]
\vspace{10.0cm}
\includegraphics{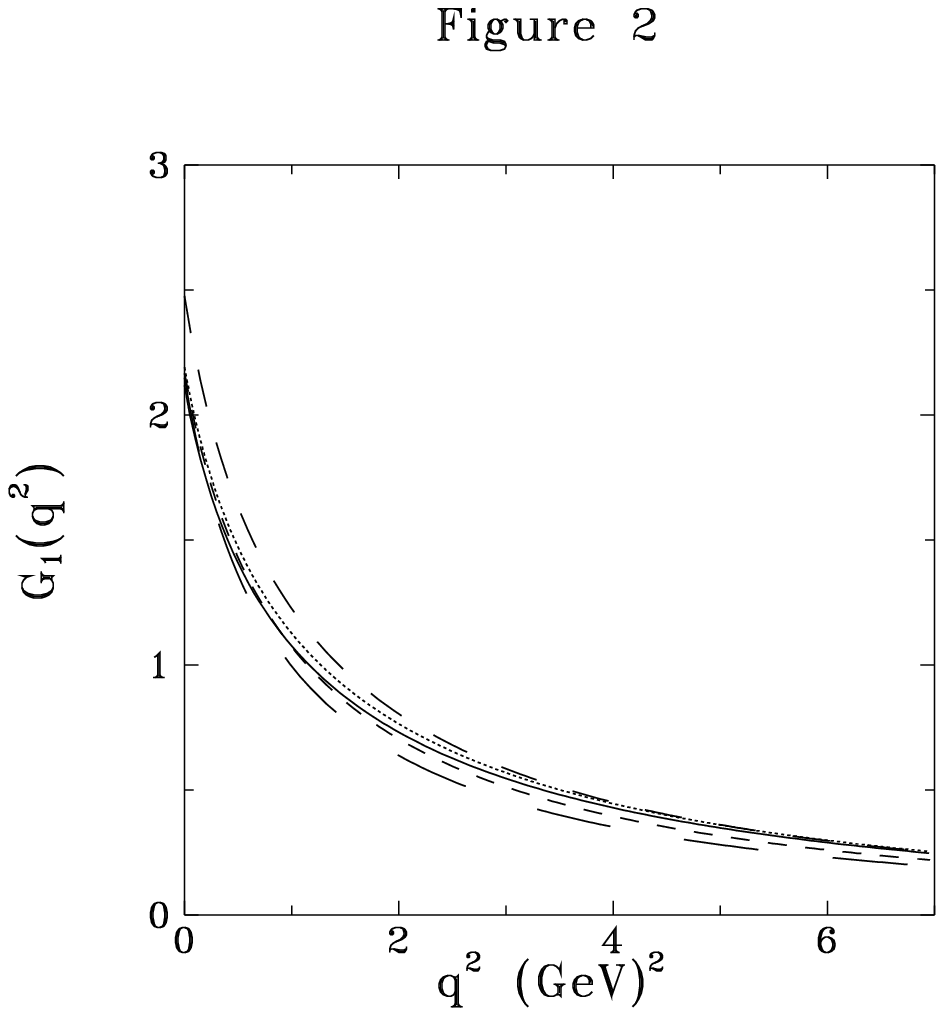}
\caption{ Magnetic form-factor $ G_1(q^2) $ 
for the $\rho$-meson as a function of $q^2$, calculated
with covariant and light-front schemes.
The curves are labeled according to Fig.1 .}
\label{fig2}
\end{figure}

\newpage
.

\begin{figure}[h]    
\vspace{10.0cm}
\includegraphics{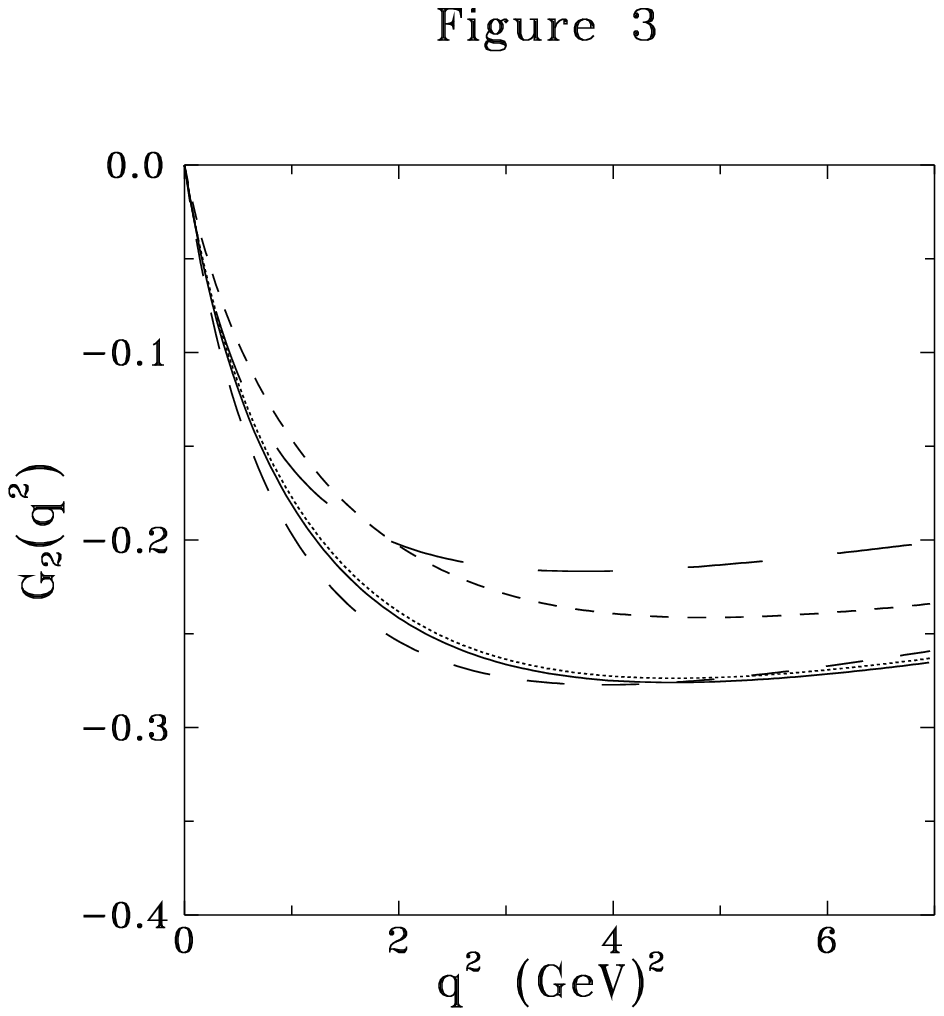}
\caption{ Magnetic form-factor $ G_2(q^2) $
for the $\rho$-meson as a function of $q^2$, calculated
with covariant and light-front schemes.
The curves are labeled according to Fig.1 .}
\label{fig3}
\end{figure}

\end{document}